\documentclass[12pt]{article}
\usepackage[OE]{express}

\usepackage{graphicx}
\usepackage{amsmath, amssymb}
\usepackage{bm}

\usepackage{multirow} 
\usepackage{booktabs}

\renewcommand{\Re}{\mathop{\mathrm{Re}}\nolimits}
\renewcommand{\Im}{\mathop{\mathrm{Im}}\nolimits}

\newcommand{\TE}{{\rm TE}}
\newcommand{\TM}{{\rm TM}}

\newcommand\cc{\mathrm{c}}
\newcommand\ee{\mathrm{e}}
\newcommand\dd{\, \mathrm{d}}
\newcommand\ii{\mathrm{i}}
\newcommand\omegap{\omega_{\rm p}}

\newcommand\lambdap{\lambda_{\rm p}}

\newcommand{\eps}{\varepsilon}
\newcommand{\lt}{\left}
\newcommand{\rt}{\right}

\newcommand{\mat}{\bm}

\newcommand\nm{\textrm{~nm}}

\newcommand\sinv{\textrm{~s}^{-1}}
\newcommand\s{\textrm{~s}}

\newcommand\Reals{\mathbb R}
\newcommand\Complex{\mathbb C}

\newcommand\diff[2]{\frac{\dd #1}{\dd #2}}


\begin{document}

\noindent{\small
This is a translation of the following paper:
}

\hspace{0.01\textwidth}
\parbox{0.99\textwidth}
{
\small
D. A. Bykov and L. L. Doskolovich, ``On the use of the Fourier modal method for calculation of localized eigenmodes of integrated optical resonators,'' {Computer Optics} {39}, 663--673 (2015).
}

\vspace{0.5em}
\noindent{\small
BiBTeX citation:
}\vspace{0.25em}

{
\small
\noindent\hspace{0.02\textwidth}
\parbox{0.99\textwidth}{\texttt{@article\{Bykov:2015:co,}}\vspace{0.3em}

\noindent\hspace{0.05\textwidth}
\parbox{0.99\textwidth}{
\texttt{title = \{\{On the use of the Fourier modal method for calculation of localized eigenmodes of integrated optical resonators\}\},\\
  author = \{Bykov, D. A. and Doskolovich, L. L.\},\\
  journal = \{Computer Optics\},\\
  volume = \{39\},\\
  number = \{5\}\,\\
  pages = \{663-{}-673\},\\
  year = \{2015\},\\
	doi=\{10.18287/0134-2452-2015-39-5-663-673\}}}\vspace{0.3em}
	
\noindent\hspace{0.02\textwidth}
\parbox{0.99\textwidth}{\texttt{\}}}
}

\vspace{0.75em}
\noindent
\hrulefill
\vspace{1em}

\title{On the use of the Fourier modal method \\
for calculation of localized eigenmodes \\
of integrated optical resonators}

\author{Dmitry A. Bykov,\!\authormark{*} and Leonid L. Doskolovich}
 
\address{Image Processing Systems Institute --- Branch of the Federal Scientific Research Centre ``Crystallography and Photonics'' of Russian Academy of Sciences, 151 Molodogvardeyskaya st., Samara 443001, Russia\\
\vspace{0.2em}
Samara National Research University, 34 Moskovskoye shosse, Samara 443086, Russia}
\email{\authormark{*}bykovd@gmail.com}

\begin{abstract}
We propose the generalization of the Fourier modal method aimed at calculating localized eigenmodes of integrated optical resonators. 
The method is based on constructing the analytic continuation of the structure's scattering matrix and calculating its poles. 
The method allows one to calculate the complex frequency of the localized mode and the corresponding field distribution. 
We use the proposed method to calculate the eigenmodes of rectangular dielectric block located on metal surface. 
We show that excitation of these modes by surface plasmon-polariton (SPP) results in resonant features in the SPP transmission spectrum. 
The proposed method can be used to design and investigate optical properties of integrated and plasmonic optical devices.
\end{abstract}


\section{\label{sec:intro}Introduction}

\begin{figure}[tbh]
    \centering
	\includegraphics{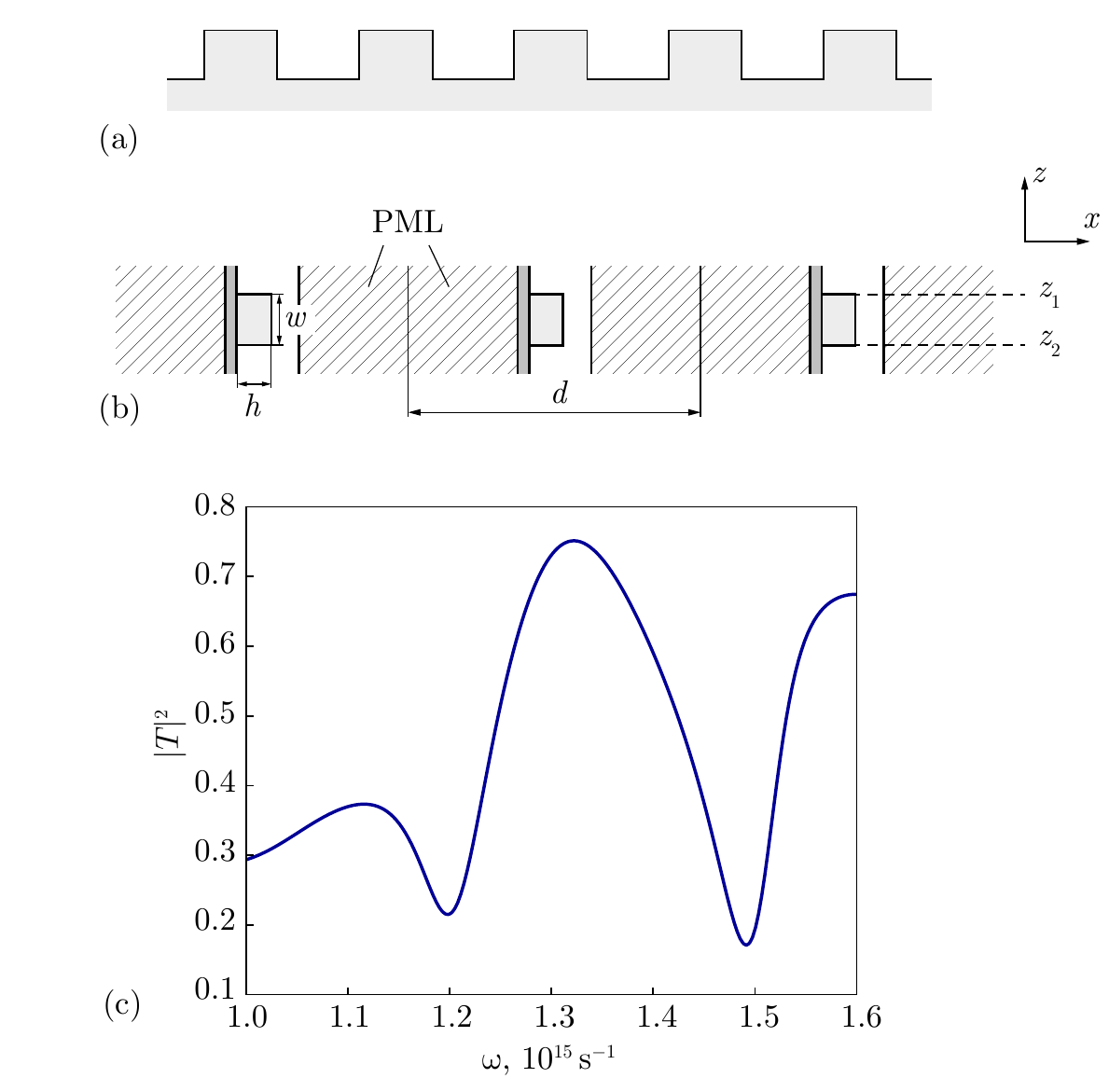}
    \caption{\label{fig1}
		(a)~Geometry of a periodic structure: an array of ridges on a metal layer (the structure is infinite in the $y$-direction);
		(b)~geometry of an aperiodic structure: 2D rectangular cavity on a thick metal layer (the structure is infinite in the $y$-direction);
		(c)~surface plasmon-polariton transmission spectrum of a rectangular cavity (length $w=900\nm$, height $h=600\nm$, permittivity $\eps=5.5$) on a silver layer with thickness $200\nm$}
\end{figure}

	In recent years, study of optical properties of resonant diffraction structures has been given considerable attention~\cite{Tikhodeev:2002:prb,Gippius:2005:prb,Weiss:2011:josaa,Belotelov:2012:josab,my:Bykov:2010:jmo,my:Bykov:2013:jlt}. 
	A subwavelength diffraction grating may serve as a simplest example of a periodic resonant diffraction structure (Fig.~\ref{fig1}a). 
	In such structures sharp resonant maxima and minima in reflection and transmission spectra are observed.
	Such resonances are caused by the excitation of the structure eigenmodes~\cite{Tikhodeev:2002:prb,my:Bykov:2010:jmo,my:Bykov:2013:jlt}.
	These modes are described by the complex-valued frequency~\cite{Tikhodeev:2002:prb,Gippius:2005:prb, my:Bykov:2013:jlt}.
	In periodic structure, modes can be either quasiguided modes, which propagate along the periodicity axis~\cite{Tikhodeev:2002:prb}, or localized modes, which are supported by the ridges or grooves of the structure~\cite{Belotelov:2012:josab}.
	The most universal approach to calculating diffraction of light by periodic structures is through the rigorous coupled-wave analysis (RCWA), which is also referred to as the Fourier modal method (FMM)~\cite{Moharam:1995:josaa}. 
	Eigenmodes of the periodic structures can be calculated using a modified RCWA approach, which is based on calculating poles of the analytic continuation of the S-matrix of the structure~\cite{Tikhodeev:2002:prb,my:Bykov:2013:jlt}.
	
	For practical purposes, of great interest are resonances that occur in \emph{aperiodic} diffraction structures. 
	Diffraction by aperiodic structures can be calculated using the Aperiodic Fourier Modal Method (AFMM)~\cite{Silberstein:2001:josaa,Hugonin:2005:josaa}.
	This method analyzes a periodic structure whose adjacent periods are optically isolated.
	The isolation is achieved either using anisotropic perfectly matched layers (PML) or gradient absorbing layers~\cite{Silberstein:2001:josaa}, or through complex coordinate transform~\cite{Hugonin:2005:josaa}.

	When dealing with \emph{aperiodic} diffraction structure, resonances similar to those observed in diffraction gratings occur, except for them being exclusively caused by the excitation of localized modes. 
	Calculation of localized modes in aperiodic structures is important when designing and investigating integrated and plasmonic optical components, such as cavities of vertical-cavity surface-emitting lasers (VCSEL) and spasers, light couplers and out-couplers. 
	Methods for calculating localized modes have been proposed in a number of papers~\cite{Mandelshtam:1997:jcp,Vallius:2006:josaa,Armaroli:2008:josaa,Weiss:2011:josaa}.
	The most popular approach uses the FDTD-based analysis of time-dependence of the electromagnetic field amplitude~\cite{Mandelshtam:1997:jcp}. 
	The major drawback of this approach is low accuracy of calculation of modes with low quality factor.
	Besides, within the FDTD-based approach it is extremely difficult to calculate the field distribution of a particular mode.
	In paper~\cite{Vallius:2006:josaa} the authors calculate the modes of laser resonator using the FMM. However, this method allow one to calculate only real-frequency modes.
	In paper~\cite{Armaroli:2008:josaa} the FMM was reformulated for cylindrical coordinate system, which allowed the authors to calculated the eigenmodes of body-of-revolution cavities.
	
	In this work, based on the aperiodic Fourier modal method, we propose a rigorous approach to calculating modes of aperiodic diffraction structures.
	The proposed method by rigorously constructing the analytic continuation of the scattering matrix allows one to calculate the complex frequency and field distribution of the modes.	
	Although the periodic and aperiodic Fourier modal methods are similar, the problem of calculating modes in the aperiodic structures is essentially more challenging, involving a number of aspects which are discussed in this paper.

\section{Modes and resonances of periodic diffraction  structures}
\label{sec:modes}

	Let us consider diffraction of the electromagnetic wave by a periodic structure (diffraction grating) shown in Fig.~\ref{fig1}a.
	Usually, the incident waves are assumed to be plane waves.
	However, in the general case, diffraction of incident periodic electromagnetic waves with period equal to that of the considered structure can also be analyzed.
	Moreover, we will consider waves that are incident from both substrate and superstrate regions.		
	
	For periodic structure, according to the Bloch--Floquet theorem, the transmitted and reflected fields can be represented as a Rayleigh expansion, i.e. an expansion into plane waves (propagating and evanescent diffraction orders).
	The same expansion can be used for the incident field.
	To solve diffraction problem we need to find amplitudes of scattered plane waves (diffraction orders) from known amplitudes of incident plane waves.
	The solution to this diffraction problem can be represented as an S-matrix~\cite{Gippius:2005:prb,my:Bykov:2013:jlt,Li:1996:josaa2}. The grating's scattering matrix $\mat{S}$ relates the complex amplitudes of incident waves ($\mat{\Psi}^{\rm inc}$) with the amplitudes of the scattered waves ($\mat{\Psi}^{\rm scatt}$) as 
\begin{equation}	 	
\label{eq1}
\mat{\Psi}^{\rm scatt} = \mat{S}\mat{\Psi}^{\rm inc},
\end{equation}
where $\mat{\Psi}^{\rm inc} = \begin{bmatrix}\mat{I}_1 \\ \mat{I}_2\end{bmatrix}$, $\mat{\Psi}^{\rm scatt} = \begin{bmatrix}\mat{R} \\ \mat{T}\end{bmatrix}$. 
	Here, $\mat{R}$ and $\mat{T}$ are the vectors composed of the complex amplitudes of the reflected and transmitted diffraction orders, while $\mat{I}_1$ and $\mat{I}_2$ are the vectors composed of the complex amplitudes of the plane waves incident on the structure from the substrate and superstrate regions. 
	An S-matrix element with indices $(i,j)$ defines the scattering amplitude of the incident wave with number $j$ into the scattered wave with number $i$.

	The S-matrix has a physical meaning only at real light frequencies.
	Let us analyze the analytic continuation of the S-matrix onto a complex-$\omega$ plane.
	Assume that the determinant of matrix $\mat S(\omega)$ has a complex pole at $\omega = \omegap$.
	In this case, $\det\mat{S}^{-1} = 0$ and there exist nontrivial solutions to the following homogeneous equation:
\begin{equation}	 	
\label{eq2}
\mat{S}^{-1}\mat{\Psi}^{\rm scatt} = \mat{0}.
\end{equation}
Thus, at $\omega=\omegap$ there is a nontrivial solution to Maxwell's equations that does not contain incident waves [see Eq.~\eqref{eq1}], which means that the structure has an eigenmode at frequency $\omegap$~\cite{my:Bykov:2013:jlt}. 
	 Now let us assume that there is an incident plane wave; let us consider the corresponding element of the S-matrix, which is a complex transmission/reflection coefficient of the structure.
	If the mode corresponding to the pole $\omegap$ of the S-matrix is excited by the considered incident wave, $\omegap$ will also be the pole of the transmission/reflection coefficient $T(\omega)$. 
	Then, the following approximate relation will hold for $T(\omega)$~\cite{my:Bykov:2012:josaa}:
\begin{equation}	 	
\label{eq3}
T(\omega) \approx a + \frac{b}{\omega-\omegap}.
\end{equation}
This equation has the meaning of a truncated Laurent series in the vicinity of point $\omega_p$.

	The transmission coefficient representation~\eqref{eq3}, which holds in the vicinity of the structure's eigenmode frequency, describes  resonant features in the transmission spectrum. 
	Thus, the problem of analyzing and understanding the resonant properties of the structure is reduced to calculating the complex frequencies of the structure's eigenmodes. 
	The real part of the complex frequency defines the resonance frequency, while the imaginary part defines the resonance quality factor $Q = \Re\omegap/(-2\Im\omegap)$.

\section{Modes and resonances of aperiodic diffraction structures}
	In this section, we analyze an aperiodic diffraction problem.
	We consider diffraction of a slab waveguide mode by a non-uniformity or resonator, located near the waveguide.
	In this case, the incident and scattered fields propagate not in a free space but in a medium, containing the waveguide.
	In particular, Fig.~\ref{fig2}b shows the considered plasmonic waveguide with rectangular block on its interface.	
	
	Analysis of aperiodic diffraction structures can be reduced to the analysis of periodic structure as follows. 
	The considered aperiodic structure is periodically continued (see Fig.~\ref{fig1}b), with the adjacent periods either being separated by perfectly matched layers (PML) (dashed area in Fig.~\ref{fig1}b)~\cite{Silberstein:2001:josaa}, or via introducing the complex coordinate transform~\cite{Hugonin:2005:josaa}. 
	This results in optical isolation of adjacent periods and, hence, the solutions to the problem of diffraction by a periodically continued structure and an aperiodic structure become identical. 
	Let us note, that instead of a homogeneous substrate (Fig.~\ref{fig1}a), one finds periodic medium above and under the structure (Fig.~\ref{fig1}b). 
	Therefore, instead of using Rayleigh expansion, the scattered and incident fields should be expanded as a sum of the modes of periodic medium.
	
	By way of illustration, consider a plasmonic mode supported by a thick silver film diffracted by a dielectric step (Fig.~\ref{fig1}b; the structure parameters are given in the figure caption). 
	The transmission spectrum is shown in Fig.~\ref{fig1}c. 
	Note that the transmission coefficient is interpreted as the ratio of the complex amplitude of the transmitted plasmonic mode to the amplitude of the incident mode. 
	The transmission spectrum has pronounced resonance features, bringing to the forefront the problem of calculating the eigenmodes (complex eigenfrequencies) of an \emph{aperiodic} structure. 
	The calculation of the eigenmodes will allow us to explain the resonant features, as well as to find numerical parameters of the resonances, such as its frequency and quality factor.
	
	For calculating poles of the S-matrix (or of transmission coefficient), one needs to be able to calculate the S-matrix (transmission coefficient) at complex values of the frequency.
	One of the most universal methods for calculating S-matrix is the Fourier modal method~\cite{Moharam:1995:josaa, my:Soifer:2014:translit}. 
	The following section deals with basic formulae of the Fourier modal method for the case of real-valued frequency.
	In Section~\ref{sec:acont} we will proceed to construct the analytic continuation.

\section{Rigorous coupled-wave analysis}
\label{sec4}

	The RCWA is based on the Bloch--Fouquet theorem, which suggests that for periodic structure, solutions to Maxwell's equations can be represented as a quasi-periodic function.
	Within the RCWA approach, the structure is assumed to consist of layers whose permittivity is $z$-independent. 
	In this case, in each layer, the field can be decomposed into a Fourier series in terms of variable $x$, which denotes the direction of structure's periodicity~\cite{Moharam:1995:josaa,my:Soifer:2014:translit} (see Fig.~\ref{fig1}a,\,b). 
	
	Let the vector $\mat\Phi_i (z)$ consist of Fourier coefficients of the electromagnetic field's tangential components in the $i$-th layer ($i=1,\ldots ,L$). 
	The vector $\mat\Phi_i (z)$ has dimension $4N$, where $N$ is the number of Fourier harmonics employed. 
	From Maxwell's equations, $\mat\Phi_i (z)$ is seen to satisfy the following matrix differential equation: 
\begin{equation} 
\label{eq4} 
\diff{\mat\Phi_i (z)}{z} = \mat A_i \cdot \mat\Phi_i (z), 
\end{equation} 
where the matrix $\mat A_i \in \Complex^{4N\times 4N}$ is defined by the $i$-th layer's geometry and the incident wave parameters (frequency and quasi-wavenumber $k_x$)~\cite{Moharam:1995:josaa,my:Soifer:2014:translit}. 
	The solution to Eq.~\eqref{eq4} describes the field in the $i$-th layer:
\begin{equation} 
\label{eq5} 
\mat\Phi_i (z)=\exp (z \mat A_i)\tilde{\mat C}_i = \mat W_i \exp (z \mat\Lambda_i ) \mat C_i,
\end{equation}
where $\mat W_i$ is a matrix with columns made up of eigenvectors of the   matrix $\mat A_i$; $\mat\Lambda_i$ is a diagonal matrix of the corresponding eigenvalues; and $\mat C_i$ is a vector of arbitrary constants.

	Relation~\eqref{eq5} should be considered as an expansion of the $i$-th layer's field in terms of the layer's eigenmodes propagating along the $z$-axis. 
	Note that the propagation constant for the $j$-th mode in the $z$-direction is derived from the eigennumber $\lambda_j$ (defining the time-dependence as $\ee^{-\ii\omega t}$ , we have $k_{z,j} =-\ii\lambda_j$).
	The transverse field distribution of the mode is defined by the $j$-th column of the matrix $\mat W_i$.
	The elements of the $j$-th column are the Fourier coefficients of the field expansion in terms of variable $x$. 
	
	To solve Maxwell's equations for a multilayered structure, we need to equate the tangential field components (or, equivalently, their Fourier coefficients) at the boundaries of the layers.
	Thus, we obtain the following set of equations: 
\begin{equation} 
\label{eq6} 
 \mat\Phi_i (z_i) = \mat\Phi_{i+1} (z_i),\qquad i=0,\ldots, L,
\end{equation} 
where $z = z_i$ is the boundary between layers $i$ and $i+1$.
System~\eqref{eq6} contains $\mat\Phi_0 (z_0)$ and $\mat\Phi_{L+1}(z_L)$, which denote the Fourier coefficients of the field at the upper and under lower boundaries the structure.   

	Let us consider the diffraction of a waveguide mode (or of a plane wave in the case of periodic structure) by the structure.
	Assume that the structure is illuminated from top by a mode whose field distribution is described by a vector of Fourier coefficients $\mat V_{\rm inc}$. 
	We assume $\mat V_{\rm inc}$ to be a column of matrix $\mat W_0$.
	Diffraction of the mode by the structure produces a set of scattered (reflected and transmitted) modes. 
	Let matrix $\mat V_r$ consist of matrix $\mat W_0$ columns that define the reflected modes and matrix $\mat V_t$ is composed of matrix $\mat W_{L+1}$ columns that define the transmitted modes. 
	Thus, the field in the superstrate region can be described as superposition of reflected modes and the incident mode:
\begin{equation}
\label{eq7} 
\mat\Phi_0 (z_0) = \mat V_r \mat R+\mat V_{\rm inc} ,\qquad \mat V_r \in \Complex^{4N\times 2N}, \mat V_{\rm inc} \in \Complex^{4N\times 1}, \mat R \in \Complex^{2N\times 1},
\end{equation} 
where $\mat R$ is the vector composed of complex amplitudes of the reflected modes. 
	The field in the substrate region is defined as superposition of transmitted modes: 
\begin{equation}
\label{eq8}
\mat\Phi _{L+1} (z_L) = \mat V_{t} \, \mat T, \qquad \mat V_t \in \Complex^{4N\times 2N}, \mat T\in \Complex^{2N\times 1},
\end{equation}
where $\mat T$ is the vector composed of complex amplitudes of the transmitted modes. 
	In the current paper we do not present the expressions for the matrices~$\mat{A},\mat{V}_r,\mat{V}_{\rm inc}, \mat{V}_t, \mat{\Phi}_i$. 
	The reader can find the general form of these matrices in~\cite{Moharam:1995:josaa, my:Soifer:2014:translit}.

	Equations~\eqref{eq6}--\eqref{eq8} form a system of linear equations in unknown amplitudes of reflected and transmitted modes ($\mat R$ and $\mat T$). 
	The diffraction problem is solved by solving this system of linear equations. 
	It is worth noting that in practice the numerical instability of the system's solution can be avoided using a variety of techniques~\cite{Moharam:1995:josaa,Li:1996:josaa2,Moharam:1995:josaa2,my:Soifer:2014:translit}.
	Besides, when calculating the matrices $\mat{A}_i$ a number of special techniques need to be used~\cite{Li:1996:josaa,my:Soifer:2014:translit}.
	
	If there is a homogeneous medium above and under the structure (see Fig.~\ref{fig1}a), the modes of the $0$-th and ($L+1$)-st layers are plane waves (propagating and evanescent diffraction orders).
	Thus, matrices $\mat A_0, \mat A_{L+1}$ take a simple form, and matrices $\mat V_t $, $\mat V_r$, $\mat V_{\rm inc}$ are defined analytically.
	In this case, expansions~\eqref{eq7} and~\eqref{eq8} are referred to as the Rayleigh expansions.
	
	In the general case, matrices $\mat V_t$, $\mat V_r$, $\mat V_{\rm inc}$ are built of the corresponding columns of matrix $\mat W$. 
	Here, the term ``corresponding'' is understood as follows. 
	Matrix $\mat V_t$ is supposed to contain matrix $\mat W_{L+1}$ columns that describe transmitted modes, matrix $\mat V_r$ contains matrix $\mat W_0$ columns that describe reflected modes, and vector $\mat V_{\rm inc}$ is matrix $\mat W_0$ column that describes the incident mode. 
	In most cases (e.g. in the case of reciprocal materials), each mode with propagation constant $k_z$ has a ``matching'' (reciprocal) mode with propagation constant $-k_z$. 
	It is worth noting that one of the two modes is incident whereas the other is scattered. 
	Thus, for the diffraction problem to be solved, all the superstrate and substrate modes need to be classified as either incident or scattered ones.
	The modes can be classified with respect to either the propagation direction (sign of the real part of $k_z$) or the evanescence direction (sign of the imaginary part of $k_z$). 
	Classification of modes into the incident and scattered ones is a fairly delicate issue, which has to be addressed with regard for the physical meaning of a diffraction problem. 
	
\begin{figure}[tbh]
    \centering
	\includegraphics[scale=0.95]{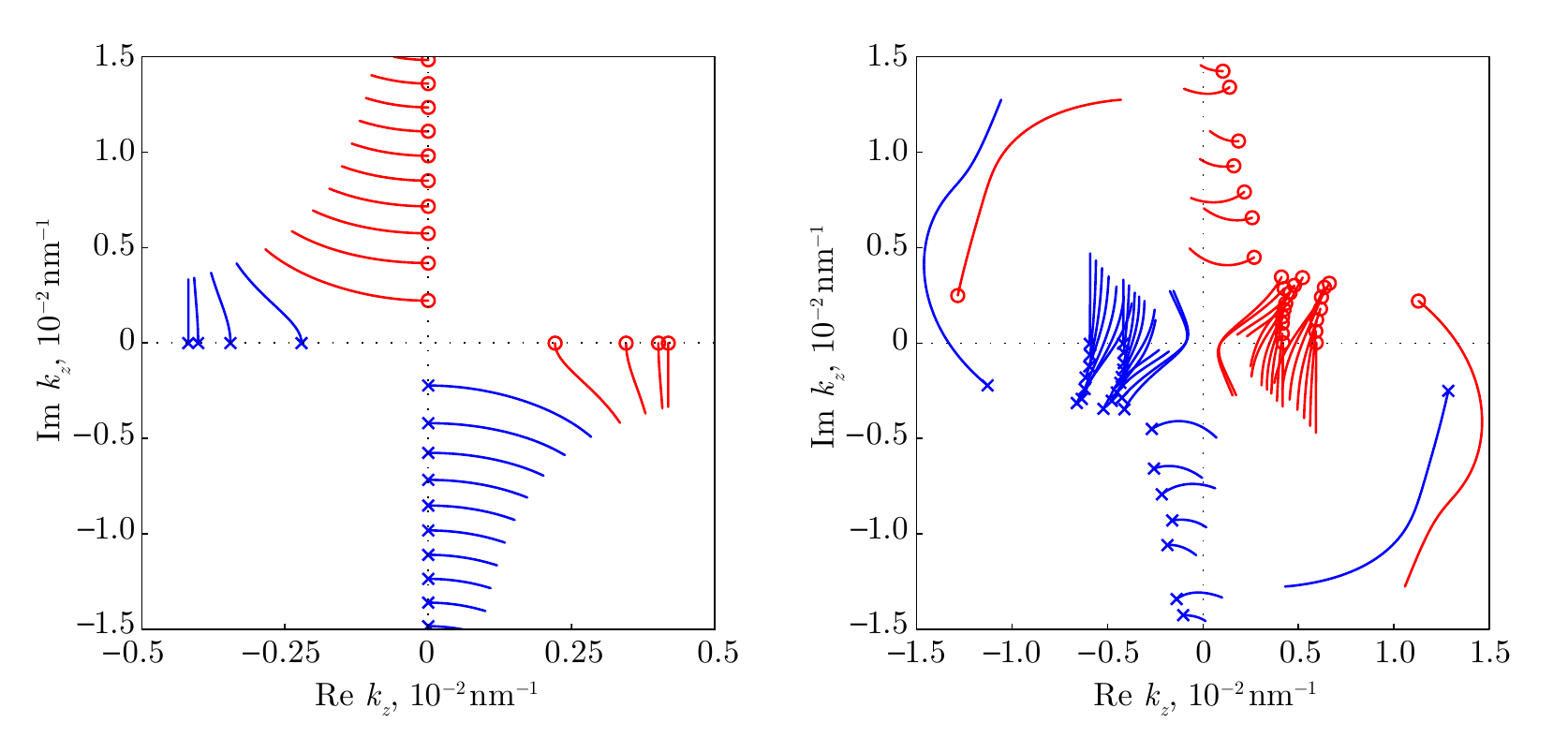}
    \caption{\label{fig2}
		Mode propagation constants $k_z$ for different superstrate geometries: 
		(a)~homogeneous medium (see Fig.~\ref{fig1}a) and 
		(b)~periodic medium (see Fig.~\ref{fig1}b; not all modes are shown). 
		For real frequency $\omega = 1.256\times10^{15}\sinv$, the incident modes are marked by crosses, scattered modes --- by circles. Curves depict the change of the propagation constants when adding a negative imaginary part ($-10^{15}\sinv\leq \Im\omega\leq 0$) to the frequency.  }
\end{figure}

	If the substrate and superstrate (see Fig.~\ref{fig1}a) are homogeneous lossless medium, there will be either propagating or evanescent modes (plane waves) above and below the structure. 
	Then, the scattered modes are chosen so that the propagating modes propagate away from the structure, whereas the evanescent ones decay with increasing distance from the structure. 
	For instance, with the time-dependence of $\ee^{-\ii\omega t} $, the superstrate scattered waves comprise propagating modes (diffraction orders) with $\Im k_z =0, \Re k_z >0$, and evanescent modes (diffraction orders) with $\Re k_z = 0, \Im k_z > 0$.
	Figure~\ref{fig2} shows the values of
$$
k_z = \pm \sqrt{\lt(\frac\omega\cc\rt)^2-\lt(\frac{2\pi}{d} n\rt)^2},
$$
where $d=5300\nm$,
for incident and scattered waves with crosses and circles, respectively.

	Given an aperiodic, rather than homogeneous, medium under the structure (an infinite grating of Fig.~\ref{fig1}b), the propagation constants $k_z$ are usually complex numbers with nonzero both real and imaginary parts. 
	Note that the direction of mode evanescence defined by the sign of $\Im k_z$ and that defined by the sign of $\Re k_z$ may differ. 
	Hence, the problem of determining the mode propagation direction needs a thorough investigation. 
	This problem becomes especially important when calculating localized modes in plasmonic elements, in which case a dielectric, a metal, and anisotropic PML-material are present both in substrate and superstrate regions.
	In this case, modes should be classified as incident and scattered as follows.
	Modes with real $k_z$ can, as before, be classified according to the propagation direction. 
	The rest modes (with complex $k_z$) should be classified according to its evanescence direction~\cite{Silberstein:2001:josaa}. 
	Hence, the reflected modes will be characterized by $\Im k_z = 0$, $\Re k_z > 0$ and $\Im k_z > 0$ (see Fig.~\ref{fig2}b).
	In this case, it is guaranteed that the solution of the diffraction problem gives field that does not increase with increasing the distance from the structure. 
	
	Thus, the diffraction problem can only be solved after accurately choosing the reflected, transmitted, and incident modes from the entire set of the substrate/superstrate modes. 
	For example, this allow one to find the mode transmission coefficient as a function of real frequency ($T(\omega)$). 
	In order to understand the mechanisms behind the resonances, in the following section we construct the analytic continuation of function $T(\omega)$ onto the complex $\omega$-plane.

\section{Constructing the analytic continuation}\label{sec:acont}

	In the case of complex frequency, similarly to the RCWA discussed above, to calculate the transmission coefficient we need to calculate eigendecomposition
of matrix $\mat A$ for the substrate/superstrate regions. 
	Then, the modes (i.e. the eigenvectors and the corresponding eigenvalues) should be classified into two groups, namely, the incident and scattered ones. 
	To make sure that $T(\omega)$ is analytic, the elements of matrices $\mat V_t$, $\mat V_r$, and $\mat V_{\rm inc}$ should be analytic functions of frequency. 
	The key difference from the real-frequency case is that with complex frequencies we are unable to rely upon physical considerations when classifying modes as the incident or scattered ones.
	
	Figure~\ref{fig2}a shows that the propagation constants for the incident (blue curves with crosses) and scattered (red curves with circles) modes above the structure change once the imaginary part of the frequency appears ($\Im\omega <0$).
	The plot suggests that once the frequency is complemented with the imaginary part, the scattered (reflected) modes, which used to be propagating ($k_z \in \Reals^{+} $), become exponentially increasing: $k_z$ acquires the negative imaginary part. Thus, if the frequency is complex, we are unable to rely upon the mode propagation or evanescence direction for the $\mat V_t$, $\mat V_r$, and $\mat V_{\rm inc}$ matrices to be constructed.
	
	In a simplest case of a homogeneous dielectric substrate/superstrate (Fig.~\ref{fig1}a), the field in the substrate/superstrate regions can be represented analytically~\cite{Moharam:1995:josaa}.
	In this case, constructing the analytic continuation of the matrices $\mat V_t$, $\mat V_r$, and $\mat V_{\rm inc}$ (and, hence, the analytic continuation of $T(\omega)$) presents a trivial problem~\cite{Tikhodeev:2002:prb}.
	
	In the general case of a nonhomogeneous periodic medium found above and under the structure (Fig.~\ref{fig1}b), the analytic relations for the eigenvectors of the matrix $\mat A$ are unknown.
	This makes the problem of constructing the analytic continuation much more challenging. 
	The only condition based on which the modes should be classified as incident/scattered (which is equivalent to constructing the above matrices) is formulated as follows. 
	If the frequency is complex, the mode is incident (scattered) if and only if with imaginary part decreasing to zero, the mode of interest becomes incident (scattered) in the sense of the real-frequency definition of the incident (scattered) mode. 
	
	Thus, the modes can be classified as the incident or scattered ones by, first, classifying them for a real frequency and, then, ``tracking'' each of the modes while gradually decreasing the (negative) imaginary part of the frequency. 
	In practice, it will suffice to ``track'' the change of the propagation constant using the following technique.  
	First, the substrate/superstrate modes need to be calculated for real frequency $\omega_0 = \Re\omega$ and classified as incident/scattered modes. 
	Then, the modes need to be calculated for complex frequency $\omega_1 = \Re\omega + \ii \Im\omega$.
	Each mode at frequency $\omega_1$ is put into one-to-one correspondence with the ``nearest'' mode at frequency $\omega_0$.
	The mode at frequency $\omega_1$ is assumed to be incident (scattered) if the corresponding mode at frequency $\omega_0$ is incident (scattered). 
	
	The proposed method allows each mode at frequency $\omega_1$ to be put in correspondence with the mode at frequency $\omega_0$, with the total change of the propagation constants being minimized. 
	Note that building a one-to-one correspondence of modes is the most time-consuming operation. 
	The following approach is proposed. 
	We build a ``proximity'' matrix with its elements defined as
\begin{equation}
\label{eq9}\lt(\mat P\rt)_{i, j} =\left|k_{z,i} -\tilde{k}_{z,j} \right|^2 , 
\end{equation} 
where $k_{z,i}$ are the propagation constants for the modes at frequency $\omega_0$ and $\tilde{k}_{z,j}$ is the set of the propagation constants at frequency $\omega_1$. 
	Then, for the cost matrix $\mat P$, an assignment problem is solved using the Hungarian algorithm~\cite{Kuhn:1955:nrlq}.
	Generally speaking, if for the mode of interest the imaginary part of frequency is large enough, even the use of the proposed algorithm can give an incorrect result because the propagation constants of the modes at frequency $\omega_1$ may be essentially different from those of the modes at frequency $\omega_0$. 
	In this case, it is advisable to utilize the proposed technique $K$ times, while consecutively increasing the imaginary part of the frequency: $\omega_k = \Re\omega+\ii \frac{k}{K}\Im\omega$, $k=0\ldots K$. 
	The ``evolution'' curves in Fig.~\ref{fig2} that illustrate the change of the propagation constant have been calculated using the above-described technique with a large value of $K$ being used. 
	To calculate the modes in the following part of the paper we use $K = 1 + \lt[-\Im\omega \cdot 10^{-13}\s\rt]$, where square brackets denote the floor operation.

	Note that in some particular cases the modes can be classified as incident/scattered ones in an approximate way~\cite{Weiss:2011:josaa}, say, based on the sign of $\Re k_z + \Im k_z$.
	However, such an approximate approach is not always suitable. 
	In particular, the classification appears to be incorrect at large values of the imaginary part of frequency (see Fig.~\ref{fig2}b). 
	Besides, an approximate classification of the modes results in incorrect calculation of the scattered field. 
	
	In this section, we have discussed constructing the analytic continuation of the transmission coefficient $T(\omega)$.
	The analytic continuation of the S-matrix is built in a similar way. 

\section{Rayleigh anomalies}

	In the previous section, we described a general approach to classification of the modes as the incident and scattered ones. 
	However, one important case has been omitted. 
	If one mode has the propagation constant $k_z=0$, two modes with $k_z=0$ are supposed to be present in the set of modes. 
	One of them should be considered as incident, while the other one as scattered. 
	Thus, in this case the incident mode is indiscernible from the scattered one. 
	
	For the diffraction grating, (Fig.~\ref{fig1}a), this special case corresponds to Rayleigh anomalies. 
	These anomalies are observed at the Rayleigh frequencies of one of diffraction orders. 
	Below the Rayleigh frequency, the diffraction order is evanescent ($\ii k_z \in \Reals$), above the Rayleigh frequency it is propagating ($k_z \in \Reals$). 
	Speaking in terms of the analytic continuation, the Rayleigh anomalies are branch points of the transmission function $T(\omega)$.  
	Presence of the branch points leads to a number of remarkable features in the transmission spectrum of diffraction gratings~\cite{my:Bykov:2012:josaa}.
	
	Similar anomalies are also observed in perfectly conducting optical elements~\cite{Kirilenko:1993:em}, in particular, when coupling perfectly conducting slab waveguides or putting inhomogeneities in perfectly conducting slab waveguides. 
	This is because the dispersion relation for modes of a slab waveguide with perfectly conducting interfaces is identical to the dispersion relation for plane waves in the Rayleigh expansion. 
	For integrated optical elements operating in the visible range, dispersion laws for the incident and scattered modes are much more complex.
	Because of this, analogs of the Rayleigh anomalies can only be observed accidentally, if two complex propagation constants of the incident and scattered waves get coincident (at a real or complex frequency). 
	It seems unlikely that such a situation may occur at practice.

\section{Calculating modes of an aperiodic structure }

	According to Eq.~\eqref{eq2}, the structure modes correspond to the poles of the S-matrix and, hence, it is most natural to find them by calculating the poles of the S-matrix of the structure. 
	In the simplest case, the calculation of the S-matrix poles is reduced to solving numerically the following equation: 
\begin{equation}
\label{eq10}
\frac{1}{\det\mat S} = 0.
\end{equation}
Note that the straightforward solution of Eq.~\eqref{eq10} leads to the numerical instability. 
	Because of this, a number of numerically stable methods have been developed for the calculation of poles of the S-matrix~\cite{my:Bykov:2013:jlt, Weiss:2011:josaa}. 
	Note that some techniques discussed in~\cite{my:Bykov:2013:jlt} require to calculate the derivative of the S-matrix, which should be done with caution: when taking the derivative using finite-difference formulae one should check that the corresponding modes in the substrate and superstrate regions are ordered in the same way at different frequencies~\cite{Weiss:2011:josaa}.
	
	The calculation of modes based on calculating the S-matrix poles is the most universal approach. 
	However, with this approach, alongside finding all localized modes of the structure, one may also derive unphysical B{\'e}renger modes localized in the PML-layers, which have no physical meaning.
	Thus, when calculating modes as poles of the S-matrix, one has to calculate the field distributions for all modes and then discard nonphysical solutions.
	If we are only interested in modes that directly affect the transmission spectrum (Fig.~\ref{fig1}c) it is advisable to calculate modes as poles of the transmission coefficient. 
	In this case, the eigenfrequencies of the modes are derived numerically as complex roots of the following equation: 
\begin{equation}
\label{eq11}
\frac{1}{T(\omega)} = 0.
\end{equation}

	By way of illustration, using the above-described technique, we calculated modes of a rectangular cavity on a metal substrate (Fig.~\ref{fig1}b). 
	The metal substrate was made of silver, with its permittivity described by the Lorentz--Drude model~\cite{Rakic:1998:ao}. 
	The cavity width was $w=900\nm$, height $h=600\nm$, permittivity of the cavity material $\eps = 5.5$, and metal layer thickness $200\nm$. 
	The modes were calculated for the frequency range $\Re\omega \in \left[1.0\times 10^{15}; 1.6\times 10^{15} \right]\sinv$. 
	Note that we analyzed only high-Q modes ($\Im\omega \in \left[-1.5\times 10^{14}; 0\right]\sinv$). 
	Frequencies of the modes were derived as roots of Eq.~\eqref{eq11}.
	For complex arguments, the transmission coefficient $T(\omega)$ was calculated using the above-discussed algorithm based on the Fourier modal methods~\cite{Moharam:1995:josaa,Li:1996:josaa2,Li:1996:josaa}.
	We used $N= 75\cdot 2 + 1$ Fourier harmonics during in the following calculations.
	
\begin{figure}[th]
	\centering
		\includegraphics{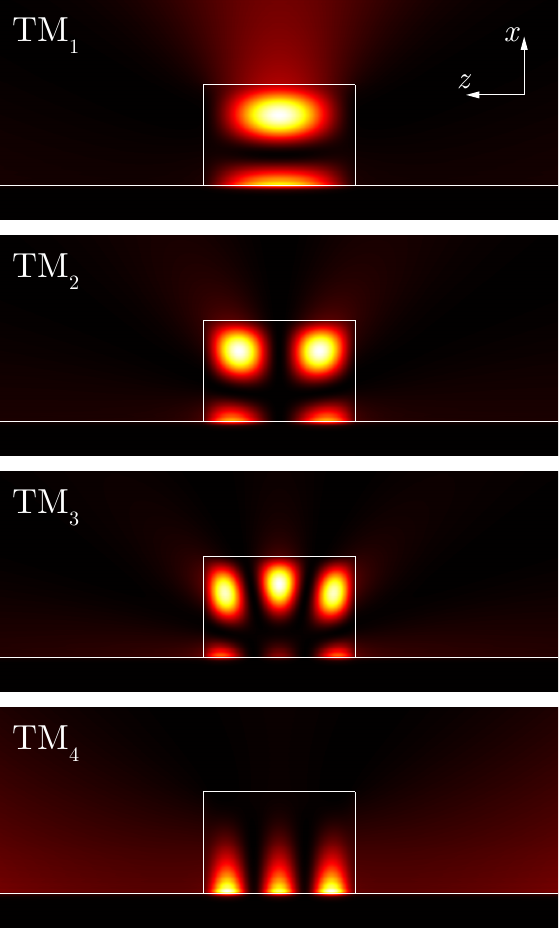}
	\caption{Field distribution ($|H_y|^2$) of TM-modes $\TM_1, \TM_2, \TM_3, \TM_4$}
	\label{fig3}
\end{figure}

	Figure~\ref{fig3} shows the electromagnetic field distribution of the modes, which was derived using a numerically stable method~\cite{Vallius:2006:josaa}.
	Corresponding complex frequencies are given in Table~\ref{table1}.
	Note that the use of an approximate technique proposed in~\cite{Weiss:2011:josaa}, in which the modes are classified into the incident and scattered ones according to the sign of $\Re k_z + \Im k_z$, gives an incorrect field distribution, which sharply increases outside the cavity region. 
	
\begin{table}
\caption{\label{table1}Mode parameters calculated using the proposed method and using MEEP (in parentheses)}
\centering
\begin{tabular}{cccc} 
\toprule
Mode & $\omegap,\sinv$ & $\Re\lambdap,\nm$ & $Q$ \\
\midrule
\multirow{ 2}{*}{$\TM_1$} & $1.011\times 10^{15}-1.092\times 10^{14}\ii$  & 1841.3 & 4.6 \\
													& $\lt(1.007\times 10^{15}-1.092\times 10^{14}\ii\rt)$ & (1848.1) & (4.6) \\ \hline
\multirow{ 2}{*}{$\TM_2$} & $1.201\times 10^{15}-4.702\times 10^{13}\ii$  & 1565.7 & 12.8 \\
													& $\lt(1.217\times 10^{15}-4.241\times 10^{13}\ii\rt)$ & (1545.3) & (14.3) \\ \hline
\multirow{ 2}{*}{$\TM_3$} & $1.505\times 10^{15}-3.932\times 10^{13}\ii$  & 1250.8 & 19.1 \\
													& $\lt(1.503\times 10^{15}-3.871\times 10^{13}\ii\rt)$ & (1252.3) & (19.4) \\ \hline
\multirow{ 2}{*}{$\TM_4$} & $1.290\times 10^{15}-1.320\times 10^{14}\ii$  & 1444.8 & 4.9 \\
													& $\lt(1.283\times 10^{15}-1.312\times 10^{14}\ii\rt)$ & (1453.2) & (4.9) \\ \midrule
\multirow{ 2}{*}{$\TE_1$} & $1.049\times 10^{15}-8.939\times 10^{13}\ii$  & 1782.9 & 5.9 \\
													& $\lt(1.050\times 10^{15}-8.569\times 10^{13}\ii\rt)$ & (1782.3) & (6.1) \\ \hline
\multirow{ 2}{*}{$\TE_2$} & $1.068\times 10^{15}-4.222\times 10^{13}\ii$  & 1761.5 & 12.7 \\
													& $\lt(1.064\times 10^{15}-4.383\times 10^{13}\ii\rt)$ & (1767.9) & (12.2) \\ \hline
\multirow{ 2}{*}{$\TE_3$} & $1.466\times 10^{15}-7.423\times 10^{13}\ii$  & 1281.7 & 9.9 \\
													& $\lt(1.464\times 10^{15}-6.427\times 10^{13}\ii\rt)$ & (1283.9) & (11.4) \\ \hline
\multirow{ 2}{*}{$\TE_4$} & $1.241\times 10^{15}-4.208\times 10^{13}\ii$  & 1516.2 & 14.7 \\
													& $\lt(1.243\times 10^{15}-5.656\times 10^{13}\ii\rt)$ & (1512.5) & (11.0) \\ \hline
\multirow{ 2}{*}{$\TE_5$} & $1.522\times 10^{15}-2.332\times 10^{13}\ii$  & 1237.1 & 32.6 \\
													& $\lt(1.521\times 10^{15}-2.240\times 10^{13}\ii\rt)$ & (1238.2) & (34.0) \\ 
\bottomrule
\end{tabular}
\end{table}

	The transmission spectrum of Fig.~\ref{fig1}c is affected by the modes $\TM_2$ and $\TM_3$. These modes have the highest $Q$-factor and their excitation produces sharp minima in the transmission spectrum at frequencies $\omega = 1.2\times 10^{15} \sinv$ and $\omega = 1.5\times 10^{15} \sinv$ (see Fig.~\ref{fig1}c).
	
	From the field distribution, the modes  $\TM_1$, $\TM_2$, and $\TM_3$ are seen to be rectangular-cavity modes of the first, second, and third orders, respectively. 
	The mode $\TM_4$ is a plasmonic Fabry--P{\'e}rot-type mode. 
	The scattered field distribution shows that mode $\TM_1$ is scattered into a wave directed normally to the surface, whereas modes $\TM_2$ and~$\TM_3$ are scattered in several directions. 
	Mode $\TM_4$ is mostly scattered into a surface plasmon-polariton.
	Thus, the analysis of the mode field distribution makes it possible to determine directions (channels) in which the electromagnetic energy is scattered. 
	Using the reciprocity theorem, it can be shown that these are the same possible directions in which the considered mode can be excited~\cite{Liu:2008:jstqe}.
	Moreover, the FMM enables the coupling coefficient between the mode and incident plane wave to be numerically calculated. 
	This analysis is important when designing nanophotonic elements to perform high-efficiency excitation of waveguide modes and surface plasmon-polaritons~\cite{Liu:2008:jstqe}.
	In particular, the method for calculating modes proposed in this work was applied by the present authors in combination with the reciprocity theorem to investigate the controlled excitation of surface plasmon-polaritons in rectangular cavities made of a magneto-optical material~\cite{my:Bykov:2014:jopt}.
	
\begin{figure}[th]
	\centering
		\includegraphics{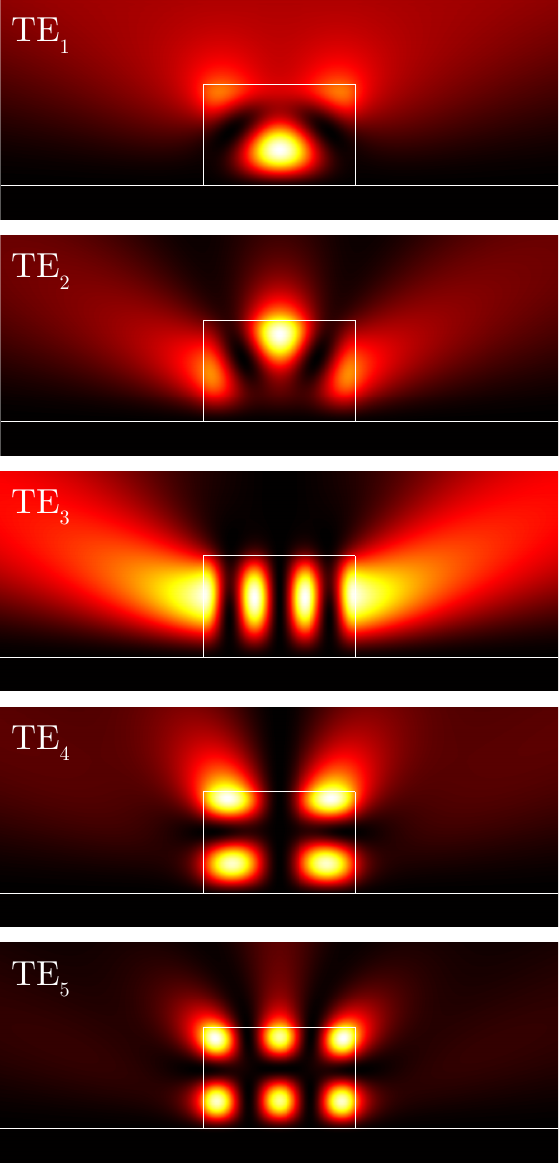}
	\caption{Field distribution ($|E_y|^2$) of TE-modes $\TE_1, \TE_2, \TE_3, \TE_4, \TE_5$}
	\label{fig4}
\end{figure}

	As we mentioned above, the calculation of the modes based on Eq.~\eqref{eq11} enables obtaining only the modes that can be excited by the considered incident field. 
	With a surface plasmon-polariton used in this work as the incident wave, the set of the calculated modes is restricted just to TM-modes.
	In addition to polarization-related limitations, the proposed approach may impose limitations on the symmetry  of the calculated modes~\cite{Gippius:2005:prb, my:Bykov:2010:jmo}. 
	To calculate all structure modes within the considered frequency range, we calculated the modes as poles of the S-matrix~\cite{my:Bykov:2013:jlt}. 
	Based on this approach, all modes of Fig.~\ref{fig3} and TE-modes of Fig.~\ref{fig4} were derived.
	The field distribution of the TE-modes shows them to be uncoupled with the surface plasmon-polariton; these modes scatter into free space above the structure.  

Table~\ref{table1} shows the complex frequencies, complex wavelengths, and quality factors of all calculated eigenmodes. 
As a matter of comparison, in parentheses the values obtained using the MEEP package~\cite{Oskooi:2010:cpc} and library Harminv are presented. MEEP and Harminv implement the FDTD approach and the method of paper~\cite{Mandelshtam:1997:jcp}.
The presented table confirms the validity of the proposed method.
The relative difference in frequencies does not exceed 7\%.
Let us note, that the developer of the MEEP~1.3 package suggests low accuracy when calculating the low-quality modes ($Q < 50$). 
Besides, the method of paper~\cite{Mandelshtam:1997:jcp} does not allow one to calculate the field of one particular mode, whereas within the proposed approach the field distribution is easily calculated (see Figs.~\ref{fig3},~\ref{fig4}).


\section{Conclusion}

	Summing up, we have proposed a generalized RCWA-based approach intended to calculate localized modes of integrated optical cavities. 
	The generalization consists in constructing the rigorous analytic continuation of the S-matrix, thus enabling modes in aperiodic structure to be calculated as poles of the S-matrix. It is of particular importance to build accurate analytic continuation when dealing with low-Q modes. The results of numerical simulation of localized modes in the dielectric structures on a metal substrate demonstrated  high efficiency of the proposed method.
	
	In this work, the problem of mode calculation has been reduced to calculating poles of the S-matrix. In an alternative approach suggested in Refs.~\cite{Vallius:2006:josaa, Rosenkrantz:2014:meta14} the modes were calculated on the assumption of generating the Fabry--P{\'e}rot type resonances. Note that methods discussed in Refs.~\cite{Vallius:2006:josaa, Rosenkrantz:2014:meta14} can also be generalized using the proposed method for constructing the analytic continuation of the S-matrix. 

\section{Acknowledgements}
The work was funded by the RSF grant (14-19-00796).

\bibliographystyle{osajnl}
\bibliography{AFMM_Modes}

\end{document}